\documentclass{elsart5p}

\usepackage{graphics}
\usepackage{graphicx}
\usepackage{amssymb}

\begin{document}

\begin{frontmatter}

\title{Multiband superconductivity and penetration depth in PrOs$_{4}$Sb$_{12}$}

\author[UCR]{D.~E.  MacLaughlin\corauthref{DEM}}
\ead{macl@physics.ucr.edu}
\author[UCR]{Lei Shu}
\author[LANL,JAEA]{R.~H. Heffner}
\author[SFU,CIAR]{J.~E. Sonier}
\author[SFU]{F.~D. Callaghan}
\author[LANL,TRIUMF]{G.~D. Morris}
\author[CSULA]{O.~O. Bernal}
\author[UCSD]{W.~M. Yuhasz}
\author[UCSD]{N.~A. Frederick}
\author[UCSD]{M.~B. Maple}

\address[UCR]{Department of Physics and Astronomy, University of California, Riverside, CA, USA}
\address[LANL]{MST-10, Los Alamos National Laboratory, Los Alamos, NM, USA}
\address[JAEA]{Advanced Science Research Center, Japan Atomic Energy Agency, Tokai, Japan}
\address[SFU]{Department of  Physics, Simon Fraser University, Burnaby, BC, Canada}
\address[CIAR]{Canadian Institute for Advanced Research, Toronto, ON, Canada}
\address[TRIUMF]{TRIUMF, Vancouver, BC, Canada}
\address[CSULA]{Department of Physics and Astronomy, California State University, Los Angeles, CA, USA}
\address[UCSD]{Department of Physics and Institute for Pure and Applied Physical Sciences, University of California, San Diego, La Jolla, CA, USA}

\corauth[DEM]{Corresponding author.}

\begin{abstract}
The effective superconducting penetration depth measured in the vortex state of PrOs$_4$Sb$_{12}$ using transverse-field muon spin rotation (TF-$\mu$SR) exhibits an activated temperature dependence at low temperatures, consistent with a nonzero gap for quasiparticle excitations. In contrast, Meissner-state radiofrequency (rf) inductive measurements of the penetration depth yield a $T^2$ temperature dependence, suggestive of point nodes in the gap. A scenario based on the recent discovery of extreme two-band superconductivity in PrOs$_4$Sb$_{12}$ is proposed to resolve this difference. In this picture a large difference between large- and small-gap coherence lengths renders the field distribution in the vortex state controlled mainly by supercurrents from a fully-gapped large-gap band. In zero field all bands contribute, yielding a stronger temperature dependence to the rf inductive measurements.
\end{abstract}


\end{frontmatter}

The superfluid density~$\rho_{\mathrm{s}}$, a fundamental property of the superconducting state, is related to the superconducting penetration depth~$\lambda$: $\rho_{\mathrm{s}} \propto \lambda^{-2}$. Thermal excitations reduce $\rho_{\mathrm{s}}$ and increase $\lambda$, with the result that the temperature dependence of $\lambda$ is an important probe of elementary excitations in the superconducting state. A fully-gapped Fermi surface results in activated behavior of the difference~$\Delta\rho_{\mathrm{s}} = \rho_{\mathrm{s}}(T) - \rho_{\mathrm{s}}(0)$ at low temperatures.
Nodes in the gap on the Fermi surface increase the density of low-lying excitations, which yields a more rapid power-law temperature dependence of $\Delta\rho_{\mathrm{s}}$. 

Transverse-field muon spin rotation (TF-$\mu$SR) experiments~\cite{MSHB02,SMHC06} in the vortex state of the filled-skutterudite heavy-fermion superconductor~PrOs$_4$Sb$_{12}$~\cite{MHYH07,ATSK07} found evidence for a fully-gapped `nodeless' superconducting state. In contrast, radiofrequency (rf) inductive measurements in the Meissner state~\cite{CSSS03} found $\Delta\lambda = \lambda(T) - \lambda(0) \propto T^2$, suggesting point nodes of the energy gap. The present paper discusses a possible resolution of this discrepancy (see also Ref.~\cite{SBMB06}).

TF-$\mu$SR experiments on PrOs$_4$Sb$_{12}$ were carried out at the M15 beam line at TRIUMF, Vancouver, Canada. Samples and experimental details have been described previously~\cite{MSHB02,SMHC06}. TF-$\mu$SR in the superconducting vortex state yields the inhomogeneous distribution of muon spin precession frequencies, i.e., the inhomogeneous field distribution, in the vortex lattice. This distribution depends on an effective penetration depth~$\lambda_{\mathrm{eff}}$ that can be estimated from rough measures of the distribution width, such as the Gaussian relaxation rate measured in the time domain, or obtained more accurately from fits to Ginzburg-Landau (\mbox{GL}) models of the distribution shape~\cite{SBK00}.

\begin{figure}[t]
\begin{center}
\includegraphics[width=0.5\textwidth]{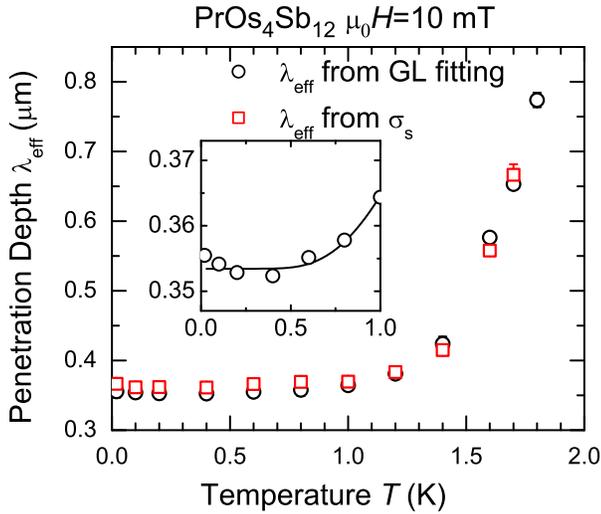}
\end{center}
\caption{Temperature dependence of effective penetration depth~$\lambda_{\mathrm{eff}}$ from vortex-state TF-$\mu$SR in PrOs$_4$Sb$_{12}$ ($T_{\mathrm{c}} = 1.8$ K). Circles: values from fits to GL model distributions (Ref.~\protect\cite{SMHC06}). Squares: values from vortex-state Gaussian relaxation rates~$\sigma_{\mathrm{s}}$.} \label{fig1}
\end{figure}

\begin{figure}[t]
\begin{center}
\includegraphics[width=0.5\textwidth]{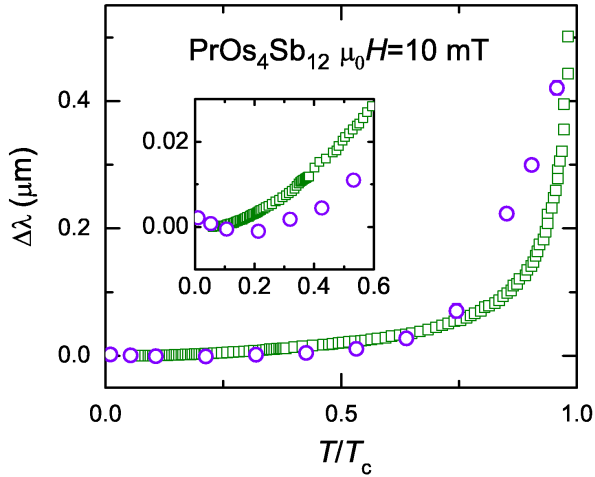}
\end{center}
\caption{Temperature dependence of $\Delta\lambda = \lambda(T) - \lambda(0)$ in PrOs$_4$Sb$_{12}$ from zero-field rf measurements (squares, Ref.~\protect\cite{CSSS03}) and vortex-state TF-$\mu$SR (circles). From Ref.~\protect\cite{SMHC06}.} \label{fig2}
\end{figure}

Figure~\ref{fig1} gives the temperature dependence of $\lambda_{\mathrm{eff}}$.
Values 
from GL model fits~\cite{SMHC06} and from Gaussian relaxation rates are in good agreement. Little temperature dependence is observed at low $T$, and the BCS form is a good fit to $\lambda_\mathrm{eff}(T)$ below $\sim$$T_{\mathrm{c}}/2$ (inset to Fig.~\ref{fig1}). A fully-gapped Fermi surface is found, with a zero-temperature gap $\Delta(0) \approx 2.2\,k_\mathrm{B}T_{\mathrm{c}}$~\cite{MSHB02}. Figure~\ref{fig2}, from Ref.~\cite{SMHC06}, shows a clear difference between the temperature dependence of $\Delta\lambda$ obtained from rf measurements~\cite{CSSS03} and from TF-$\mu$SR.

This discrepancy can be understood qualitatively in the extreme two-band scenario of Seyfarth et al.~\cite{SBMB06,SBMF05} for PrOs$_4$Sb$_{12}$, in which thermal conductivity and other data are explained by large and small gaps $\Delta_{\mathrm{L}}$, $\Delta_\mathrm{S}$ on different sheets of the Fermi surface. With band-specific Fermi velocities $v_{\mathrm{FL}}$, $v_{\mathrm{FS}}$, coherence lengths~$\xi_{\mathrm{L},\mathrm{S}} \approx \hbar v_{{\mathrm{FL}},\mathrm{S}} /\Delta_{\mathrm{L},\mathrm{S}}$ can be defined; typically $\xi_\mathrm{L} < \xi_{\mathrm{S}}$. The vortex state is then characterized by a crossover field~$H_{\mathrm{c}2}^{\mathrm{S}} = \Phi_0/2\pi\xi_{\mathrm{S}}^2 < H_{\mathrm{c}2}$, where $\Phi_0$ is the flux quantum. If the large-gap band is also a heavy band ($m_\mathrm{L} > m_{\mathrm{S}}$), then $v_{\mathrm{FL}} < v_{\mathrm{FS}}$, $\xi_\mathrm{L} \ll \xi_{\mathrm{S}}$, and $H_{\mathrm{c}2}^{\mathrm{S}}$ can $\mathrm{be} \ll H_{\mathrm{c}2}$. In PrOs$_4$Sb$_{12}$ at low temperatures $H_{\mathrm{c}2}^{\mathrm{S}} \sim 100\ \mathrm{Oe} \approx H_{c1}$~\cite{SBMB06}.

For~$H \gtrsim H_{\mathrm{c}2}^{\mathrm{S}}$ small-band vortex core states with size scale~$\xi_{\mathrm{S}}$ overlap. In PrOs$_4$Sb$_{12}$ this applies for essentially the entire vortex state, and the observed anomalous thermal conductivity~\cite{SBMB06,SBMF05} is mainly due to heat transfer by small-band excitations. Then the small-gap states and their contributions to screening supercurrents are nearly uniform, and the vortex-state field inhomogeneity is mainly due to large-gap supercurrents. The activated temperature dependence of $\lambda_{\mathrm{eff}}$ (Fig.~\ref{fig1}) is evidence that the large gap is nodeless, which is corroborated by thermal conductivity experiments in very clean single crystals~\cite{SBMB06}. In this picture TF-$\mu$SR measurements are insensitive to the nodal structure of the small gap. 

In contrast, the Meissner-state penetration depth~$\lambda$ contains contributions from both bands, and its temperature dependence is controlled by both small- and large-gap superfluid densities. At low temperatures the small-gap contribution dominates the temperature dependence, and $\lambda$ varies more rapidly than $\lambda_{\mathrm{eff}}$ as observed (inset to Fig.~\ref{fig2}). The behavior of the data at higher temperatures is more complicated and will not be discussed here. The similar discrepancy found in Sr$_2$RuO$_4$~\cite{SMHC06,MaMa03} might also be explained by multiband superconductivity in that compound.

This picture is qualitative and somewhat speculative; its chief merit is that it accounts for a number of different experimental results in PrOs$_4$Sb$_{12}$. To our knowledge there is at present no theory for the temperature dependence of the vortex-state field distribution in an extreme two-band superconductor.

\section*{Acknowledgement}
We are grateful to the TRIUMF $\mu$SR staff for their technical support during the measurements. This work was supported in part by the U.S. NSF, grant nos. 0102293 and 0422674 (Riverside), 0203524 and 0604015 (Los Angeles), and 0335173 (San Diego), by the U.S. DOE, contract DE-FG-02-04ER46105 (San Diego), and by the Canadian NSERC and CIAR (Burnaby). Work at Los Alamos was performed under the auspices of the U.S. DOE.

\end{document}